\documentclass[%
 reprint,
superscriptaddress,
 amsmath,amssymb,
 aps,
 prl
]{revtex4-2}

\usepackage{graphicx}
\usepackage{dcolumn}
\usepackage{bm,upgreek}
\usepackage{soul}
\usepackage[normalem]{ulem}
\usepackage{enumitem}
\usepackage{dsfont}
\usepackage{yhmath}
\usepackage{nicefrac}

\usepackage{color}
\usepackage{physics}
\usepackage{hyperref}

\usepackage{bibunits}

\begin{document}

\title{Breaking of reciprocity and the Pancharatnam-Berry phase for light scattered by a disordered cold atom cloud}

\author{P. H. N. Magnani}
\affiliation{Departamento de F\'{\i}sica, Universidade Federal de S\~{a}o Carlos, Rodovia Washington Lu\'{\i}s, km 235 - SP-310, 13565-905 S\~{a}o Carlos, SP, Brazil}
\author{P. G. S. Dias}
\affiliation{Departamento de F\'{\i}sica, Universidade Federal de S\~{a}o Carlos, Rodovia Washington Lu\'{\i}s, km 235 - SP-310, 13565-905 S\~{a}o Carlos, SP, Brazil}
\author{M. Frometa}
\affiliation{Instituto de F\'isica de S\~{a}o Carlos, Universidade de S\~{a}o Paulo - 13566-590 S\~{a}o Carlos, SP, Brazil}
\author{M. A. Martins}
\affiliation{Departamento de F\'{\i}sica, Universidade Federal de S\~{a}o Carlos, Rodovia Washington Lu\'{\i}s, km 235 - SP-310, 13565-905 S\~{a}o Carlos, SP, Brazil}
\author{N. Piovella}
\affiliation{Dipartimento di Fisica "Aldo Pontremoli", Universit\`{a} degli Studi di Milano, Via Celoria 16, I-20133 Milano, Italy \&
INFN Sezione di Milano, Via Celoria 16, I-20133 Milano, Italy}
\author{R. Kaiser}
\affiliation{Universit\'e C\^ote d'Azur, CNRS, INPHYNI, France}
\author{Ph. W. Courteille}
\affiliation{Instituto de F\'isica de S\~{a}o Carlos, Universidade de S\~{a}o Paulo - 13566-590 S\~{a}o Carlos, SP, Brazil}
\author{M. Hugbart}
\affiliation{Universit\'e C\^ote d'Azur, CNRS, INPHYNI, France}
\author{R. Bachelard}
\affiliation{Departamento de F\'{\i}sica, Universidade Federal de S\~{a}o Carlos, Rodovia Washington Lu\'{\i}s, km 235 - SP-310, 13565-905 S\~{a}o Carlos, SP, Brazil}
\affiliation{Universit\'e C\^ote d'Azur, CNRS, INPHYNI, France}
\author{R. C. Teixeira}
\affiliation{Departamento de F\'{\i}sica, Universidade Federal de S\~{a}o Carlos, Rodovia Washington Lu\'{\i}s, km 235 - SP-310, 13565-905 S\~{a}o Carlos, SP, Brazil}

\date{\today}

\begin{abstract}
Collective effects on the light scattered by disordered media such as Anderson localization and coherent backscattering critically depend on the reciprocity between interfering optical paths. In this work, we explore the breaking of reciprocity for the light scattered by a disordered cold atom setup, taking advantage of the non-commutation of optical elements that manipulate the polarization of the interfering paths. This breaking of symmetry manifests itself in the reduction of the fringes contrast as the light scattered by the cloud interferes with that from its mirror image. We provide a geometrical interpretation in terms of the Pancharatnam-Berry phase, which we directly access from the fringes displacement. Our work paves the way toward the manipulation of path reciprocity and interference for light scattered by disordered media.
\end{abstract}

\maketitle

{\em Introduction.---} Symmetries are fundamental properties of a system, which determine the conserved quantities of its dynamics~\cite{Noether1918,Noether1971}, thus setting constraints on its evolution. In particular, the charge, parity and time reversal (CPT) symmetry is universal as it applies to all known forces. Yet an important aspect of this symmetry is that it holds for a system as a whole, and considering the full symmetry, rather than a single one of the three. A simple illustration can be found in classical optics: While the CPT symmetry holds for the scattering of light by particles if one monitors all degrees of freedom, it breaks down as one focuses, for example, on the coherently scattered light. Indeed, the scattering of electromagnetic energy in other modes is typically encapsulated in the imaginary part of the medium refractive index, thus inducing a dissipative nature to the system. The necessity to differentiate absorption, which as a matter of fact defines an arrow for time and prevents the time reversal symmetry, from other energy-preserving mechanisms which may break that symmetry, has led to the notion of reciprocity~\cite{Potton2004,Carminati2000}.

Reciprocity, in optics and beyond, describes the similarity of the transformations undergone by two waves travelling along the same path, yet in opposite directions. The fields emerging from these two ``reciprocal paths'' can be made to interfere, and in disordered systems the constructive interference between such pairs of optical paths is at the origin of modifications to the classical diffusion of waves, such as the coherent backscattering effect~\cite{Kuga1984,Wolf1985,Albada1985,Yoo1990,Mishchenko1993,Tourin1997,Labeyrie1999} and the Anderson localization of waves~\cite{Anderson1958,Akkermans2007}. The vectorial nature of electromagnetic waves may actually lead to a reduction of the enhancement of the backscattered signal, as (de)polarization effects kick in~\cite{Kuga1984,Albada1985,Wolf1985,Akkermans1986,Stephen1986}. Similarly, the presence of multiple, coupled polarization channels for light has been reported to be detrimental to Anderson localization~\cite{Skipetrov2014,Maximo2015}, superradiance~\cite{Gross1982} and subradiance~\cite{Cipris2021}, or to the build up of a large refractive index in dense disordered media~\cite{Andreoli2021}.

\begin{figure}[h]
	\centering	\includegraphics[width=\columnwidth]{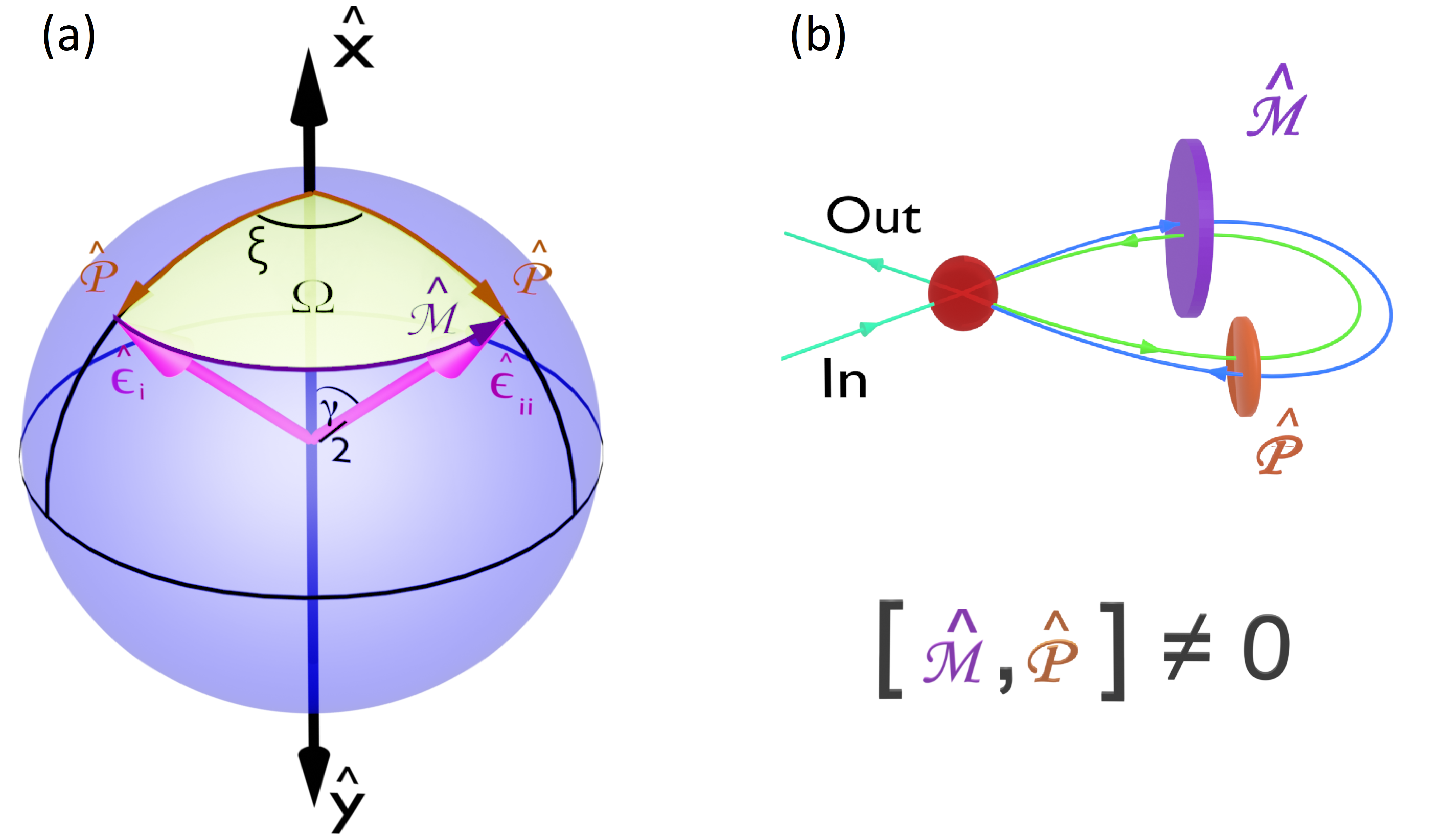}
	\caption{(a) Representation on the Poincaré sphere of the light polarization $\hat\epsilon_{i}$ and $\hat\epsilon_{ii}$, obtained after the action of the transformations $\hat{\mathcal{M}}$ and $\hat{\mathcal{P}}$ in direct or reverse order, respectively. The angles $\xi$ and $\gamma$ are defined by the operations $\hat{\mathcal{M}}$ and $\hat{\mathcal{P}}$, respectively, and the solid angle $\Omega$ is indicated in light yellow. (b) Schematic representation of the breaking of reciprocity between time-reversed paths, composed by non-commuting polarization operators $\hat{\mathcal{M}}$ and $\hat{\mathcal{P}}$.}
	\label{fig:physics}
\end{figure}

In this work, we explore the breaking of reciprocity for light scattered by a disordered medium by introducing non-commuting polarizing optics on the path of the light. The light scattered by a large cold atom cloud is made to interfere with that from the cloud mirror image, yet the use of a birefringent mirror (M), along with a half waveplate (P), breaks the symmetry in the transformation that the polarization undergoes on each (reciprocal) path. The reciprocity breaking is monitored through the interference fringes from our mirror-assisted backscattering setup \cite{Moriya16}, and the reduction of their contrast is shown to derive directly from the lack of commutation of the above-mentioned polarizing optics. While this contrast is associated to the geodesic distance between the polarizations on the Poincaré sphere, the phase difference between them, also known as the Pancharatnam-Berry phase \cite{Pancharatnam1956,Berry1987}, corresponds to the geodesic surface between the injected polarization and the scattered ones, as shown in Fig.~\ref{fig:physics}: It is accessed in our experiment by measuring the fringes displacement, thus providing a geometrical interpretation to the breaking of reciprocity.

\begin{figure}[h]
\centering
\includegraphics[width=\columnwidth]{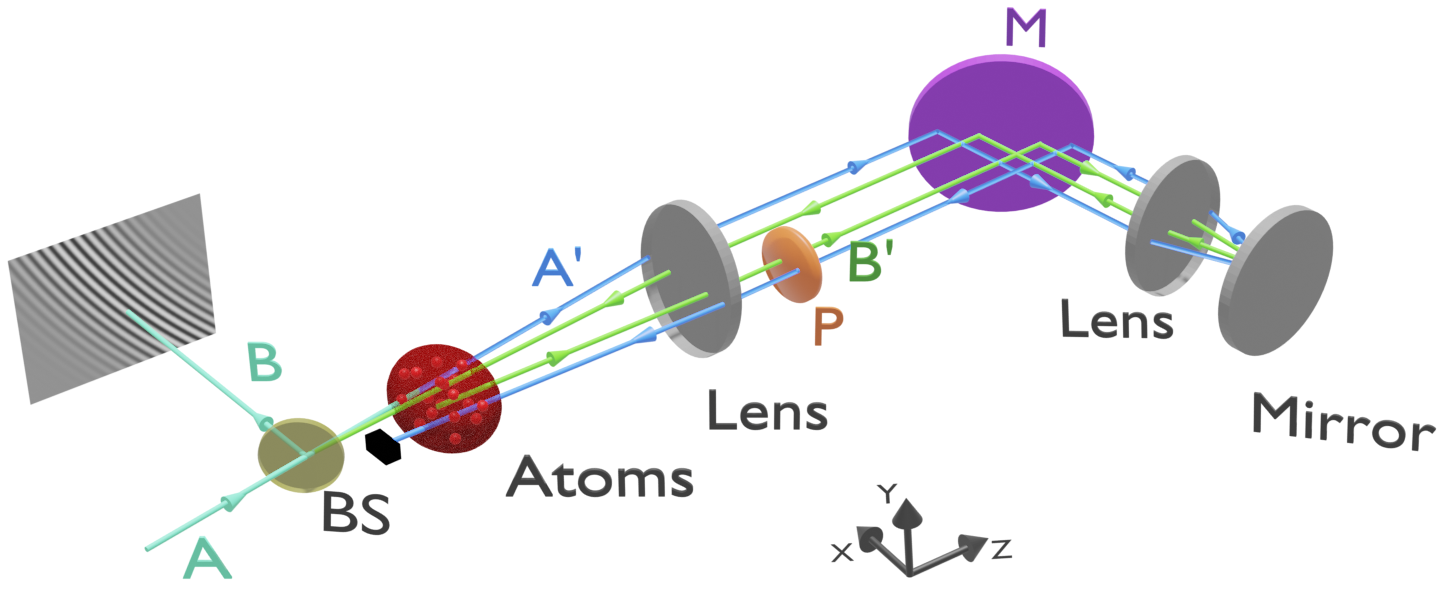}
\caption{Experimental setup. The laser light, resonant with the atomic sample, passes through a non polarising beam splitter (BS) before reaching the atomic cloud, which corresponds to path A indicated by a blue arrow. The transmitted beam passes through the 1:1 telescope composed of two lenses of same focal length, is reflected at the 45$^{\circ}$ mirror M, before being reflected back by a mirror with a small angle $\theta_0$, passing again through a lens, the 45$^{\circ}$ mirror M, and this time by a half-waveplate P before it reaches the other lens and then atoms; this stands for path A' (in blue). The scattered light by the atoms can propagate along the reverse path, where it passes the waveplate before the birefringent mirror and then travels back to the atomic cloud, which corresponds to path B' (in green); From the atomic cloud, the light follows the path B through the BS, to be reflected towards a CCD camera on which one can observe the fringes due to the interference of all the possible paths.}
\label{fig:setup}
\end{figure}

{\em Experimental Setup.---} Our cold atom setup consists of a cold cloud of  $N\approx 3.10^7$ atoms of $^{88}\text{Sr}$, initially trapped in a magneto-optical trap with a temperature of $10~$mK, see Fig.~\ref{fig:setup}. The experiment is performed by turning off the trap, so the atomic cloud expands for a time of flight of $200~\upmu$s and reaches a $1/\sqrt{e}$ radius of $R = 1100~\upmu$m. With an associated optical density $b_0\approx0.4$, single scattering is dominant. 
Then, the atoms are probed with a laser beam with a waist $w_0=2.1~$mm, so that the intensity is almost homogeneous throughout the cloud. The laser power $P_l=0.2~$mW, resonant with the $461~$nm broad transition of neutral Strontium ($\Gamma=2\pi \times 30.5~$MHz), corresponds to a saturation parameter of $s_0=0.07$, so that most of the light is scattered elastically. The maximum pulse duration $t_p=70~\upmu$s is chosen such as to neglect the Doppler displacement of the transition due to the acceleration of the atoms during the pulse.


Our experiment is thus set in the single- and elastic-scattering regime, so the radiation from each atom can be decomposed as four optical paths, composed of the paths $A$, $B$, $A'$ and $B'$ shown in Fig.~\ref{fig:setup}. Indeed, the light incoming on the atomic cloud may either:
\setlist{nolistsep}
\begin{enumerate}[label=(\roman*),noitemsep,leftmargin=0.5cm]
    \item follow $A+A'+B$, i.e., cross the atomic sample without being scattered, be reflected on the 45$^\circ$ mirror (M) before being reflected back toward the half-waveplate (P), and finally be scattered by the atoms and reach the CCD,\label{path:i}
    \item follow $A+B'+B$, i.e., first be scattered by the atoms toward the plate P, then reflected back toward the 45$^\circ$ mirror M and eventually propagate until the CCD,\label{path:ii}
    \item follow $A+B$, i.e., be directly backscattered by the atoms before being collected on the CCD, \label{path:iii}
    \item follow $A+A'+B'+B$, i.e., cross the science chamber without being deviated, be reflected on the 45$^\circ$ mirror M before being reflected toward the plate P, and be backscattered by the atoms to meet again the plate P, be reflected toward the 45$^\circ$ mirror M and propagate until the CCD.\label{path:iv} 
\end{enumerate}
 Here, \ref{path:i} and \ref{path:ii} are two reciprocal paths with very similar length, which interfere to build up fringes (according to the overlap between their polarization). In contrast, \ref{path:iii} and \ref{path:iv} correspond to a different optical path for each atom, thus contributing to a background light. Due to the presence of reciprocal paths, This mirror-assisted backscattering configuration~\cite{Greffet91} can be used to explore the light coherence in disordered atomic clouds~\cite{Moriya16,Dias2021}.

{\em Breaking of reciprocity.---} The polarization of the reciprocal paths is manipulated by using polarizing optics between the cloud and the backscattering mirror (Fig.~\ref{fig:setup}). The injected light has a linear polarization $\hat\epsilon_0=\hat{x}$. M is a dielectric mirror that induces a phase delay $\xi$ on the $\hat{y}$ component (relative to the $\hat{x}$ component) of the light polarization. The fast axis of P is tilted from the $\hat{x}$ axis by an angle $\gamma/2$, which is our control parameter, inducing a reflection of the polarization with respect to the fast axis. Within the Jones formalism for polarising optics ~\cite{Jones1941,Collett2009}, the matrices associated to M and P are, respectively:
\begin{equation}
\hat{\mathcal{M}} = 
\begin{bmatrix}
1 & 0  \\
0 & e^{i \xi}
\end{bmatrix} \; ; \; \hat{\mathcal{P}} = \begin{bmatrix}
\cos \gamma & \sin \gamma  \\
\sin \gamma & - \cos \gamma
\end{bmatrix},
\label{eq:Mat}
\end{equation}
in the $(\hat{x},\hat{y})$ basis. Then, the polarizations emerging from each path are given by: 
\begin{equation}
\begin{split}
&\hat\epsilon_{i} =\hat{\mathcal{P}}\hat{\mathcal{M}}\hat\epsilon_0,
\\ &\hat\epsilon_{ii} =\hat{\mathcal{M}}\hat{\mathcal{P}} \hat\epsilon_0,
\\ &\hat\epsilon_{iii} =\hat\epsilon_0,
\\ &\hat\epsilon_{iv} =\hat{\mathcal{M}}\hat{\mathcal{P}}\hat{\mathcal{P}}\hat{\mathcal{M}}\hat\epsilon_0=\hat{\mathcal{M}}^2\hat\epsilon_0.
\end{split}
\end{equation}

\noindent Thus, the polarization of the two paths providing the background light, \ref{path:iii} and \ref{path:iv}, remain along $\hat{x}$.

The observed fringes result from polarization effects and differences in optical path between the four paths. Indeed, accounting for the tilt between the incident beam after the beamsplitter (wavevector $\mathbf{k}_0=k(\sin\theta_0 \hat{y}-\cos\theta_0 \hat{z}$) and the backward direction of observation (wavevector $\mathbf{k}=k(\sin\theta \hat{y}+\cos\theta \hat{z}$), we obtain the following field scattered by atom $j$:
\begin{multline}
    \vec{E}_j=E_s \big[ e^{ik(\cos\theta-\cos\theta_0)z_j}\hat\epsilon_{i} +e^{ik(\cos\theta_0-\cos\theta)z_j}\hat\epsilon_{ii}
    \\ +e^{ik(\cos\theta_0+\cos\theta)z_j}\hat\epsilon_{iii} +e^{-ik(\cos\theta_0+\cos\theta)z_j}\hat\epsilon_{iv}\big],
    \label{eq:E_main}
\end{multline}
where $E_s$ is a prefactor which encapsulates the amplitude of the pump, the single-atom cross-section and the distance of the camera from the setup. Summing the fields from all atoms, and assuming a continuous Gaussian density and a small tilt $|\theta-\theta_0|\ll\theta_0$, we get~\cite{SM}
\begin{equation}
    I_\textrm{out}(\theta)=NI_a \left(1+\frac{1}{2} \Re[\langle\hat\epsilon_{i},\hat\epsilon_{ii}\rangle e^{i\phi}]S_{R}(\theta)\right),\label{eq:IntTheta}
\end{equation}
where the scalar product is defined as $\langle\hat\epsilon_{i},\hat\epsilon_{ii}\rangle=\hat\epsilon_{i}\cdot\hat\epsilon_{ii}^*$,  $\phi=2\theta_0kh(\theta-\theta_0)$ the angular variable describing the fringes, with $h$ the distance between the mirror and the center of the atomic cloud, and $S_{R}(\theta)=e^{-2(\theta_0kR)^2(\theta-\theta_0)^2}$ the fringes spatial envelope. Thus, while paths~\ref{path:iii} and \ref{path:iv} contribute only to the background, the reciprocal paths~\ref{path:i} and \ref{path:ii} interfere to provide fringes in an angle $|\theta-\theta_0|\lesssim 1/kR$. An example of these fringes and the surrounding background light can be observed in Fig.~\ref{fig:setup}.

{\em Non-commutation of the polarizing optics.---} As it can be seen in Eq.~\eqref{eq:IntTheta}, the fringes contrast is given by the modulus of the overlap (that is, the scalar product) between the polarizations of the reciprocal paths~\ref{path:i} and \ref{path:ii}, which is calculated as  
\begin{equation}
\begin{split}
\langle\hat\epsilon_{i},\hat\epsilon_{ii}\rangle &=\langle\hat{\mathcal{P}}\hat{\mathcal{M}}\hat\epsilon_0, \hat{\mathcal{M}}\hat{\mathcal{P}}\hat\epsilon_0\rangle
    \\ & =  1-\langle\hat{\mathcal{M}}\hat{\mathcal{P}}\hat\epsilon_0, [\hat{\mathcal{P}},\hat{\mathcal{M}}]\hat\epsilon_0\rangle,
\end{split}
\end{equation}
where $[\cdot,\cdot]$ refers to the commutator. In other words, the interference between the reciprocal paths [see Eq.~(\ref{eq:IntTheta})] is reduced when the matrices of the polarizing optics stop commuting. In the case of M and P used in our experiment, the contrast reads
\begin{equation}
\mathcal{C}= |\langle\hat\epsilon_{i},\hat\epsilon_{ii}\rangle| =\sqrt{1-\sin^2 (2\gamma)\sin^2(\xi/2)}.\label{eq:C}
\end{equation}
The delay $\xi$ is determined by the dielectric coating of M, and chosen close to $\pi$ to maximize the reduction of the contrast ($\xi=(162 \pm 10)^{\circ}$ in our setup~\cite{SM}), while $\gamma$ is used as a control parameter to explore the non-commutation between the optics, and subsequently the breaking of reciprocity between paths~\ref{path:i} and \ref{path:ii}.

\begin{figure}[h]
    \centering
    \includegraphics[width=0.45\textwidth]{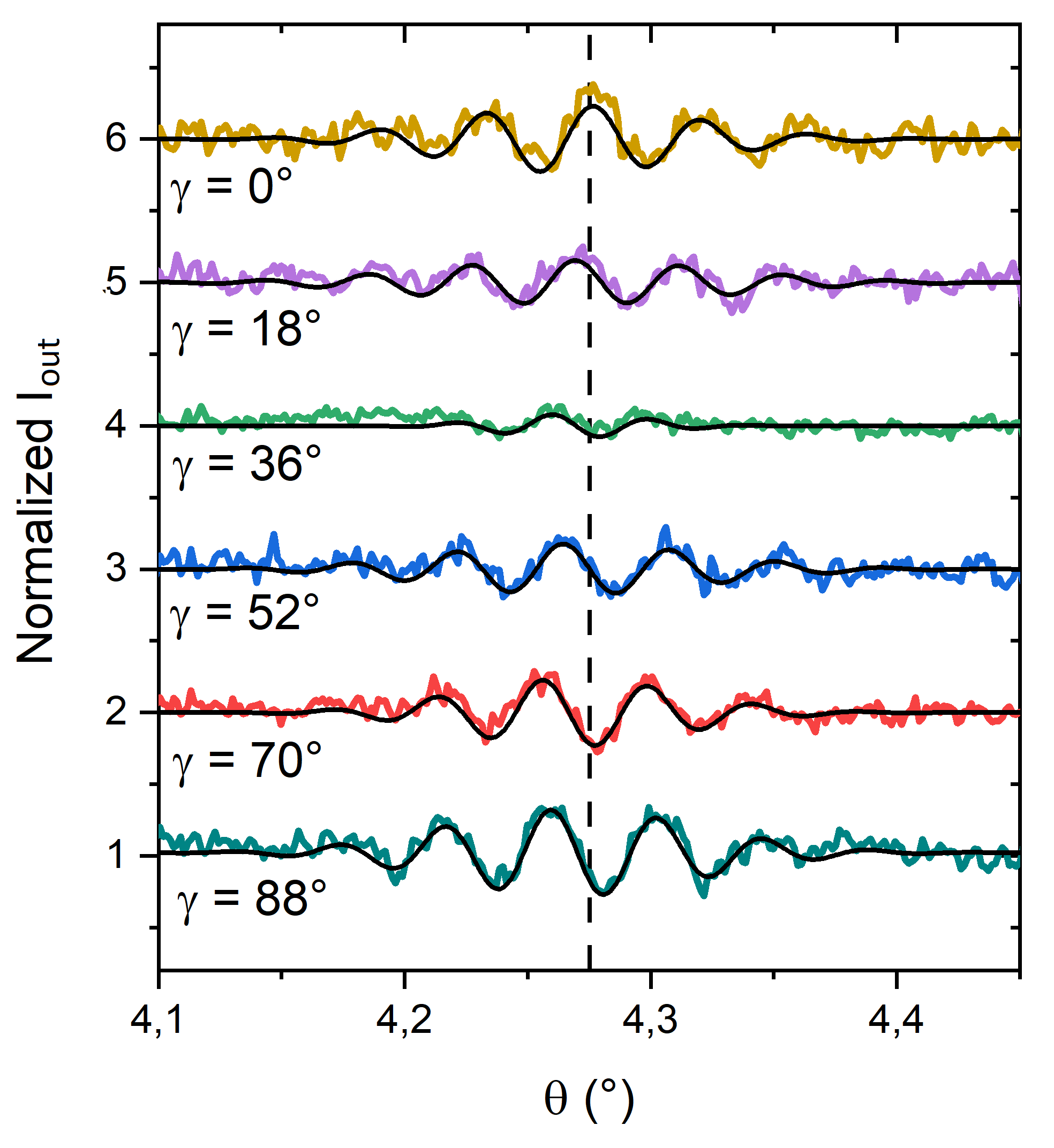}		
	\caption{Colored lines: Interference fringes integrated along the azimuthal direction, normalized to the background intensity, and shifted vertically by one for visibility. Black continuous lines: Fitted fringe profiles, following Eq.~(\ref{eq:IntTheta}). Dashed vertical line: center of the fringes' envelope, $\theta_0$, found by a global fit of the envelope to all curves.}
    \label{fig:fringes}
\end{figure}

\begin{figure}[h]
    \centering
    \includegraphics[width=0.45\textwidth]{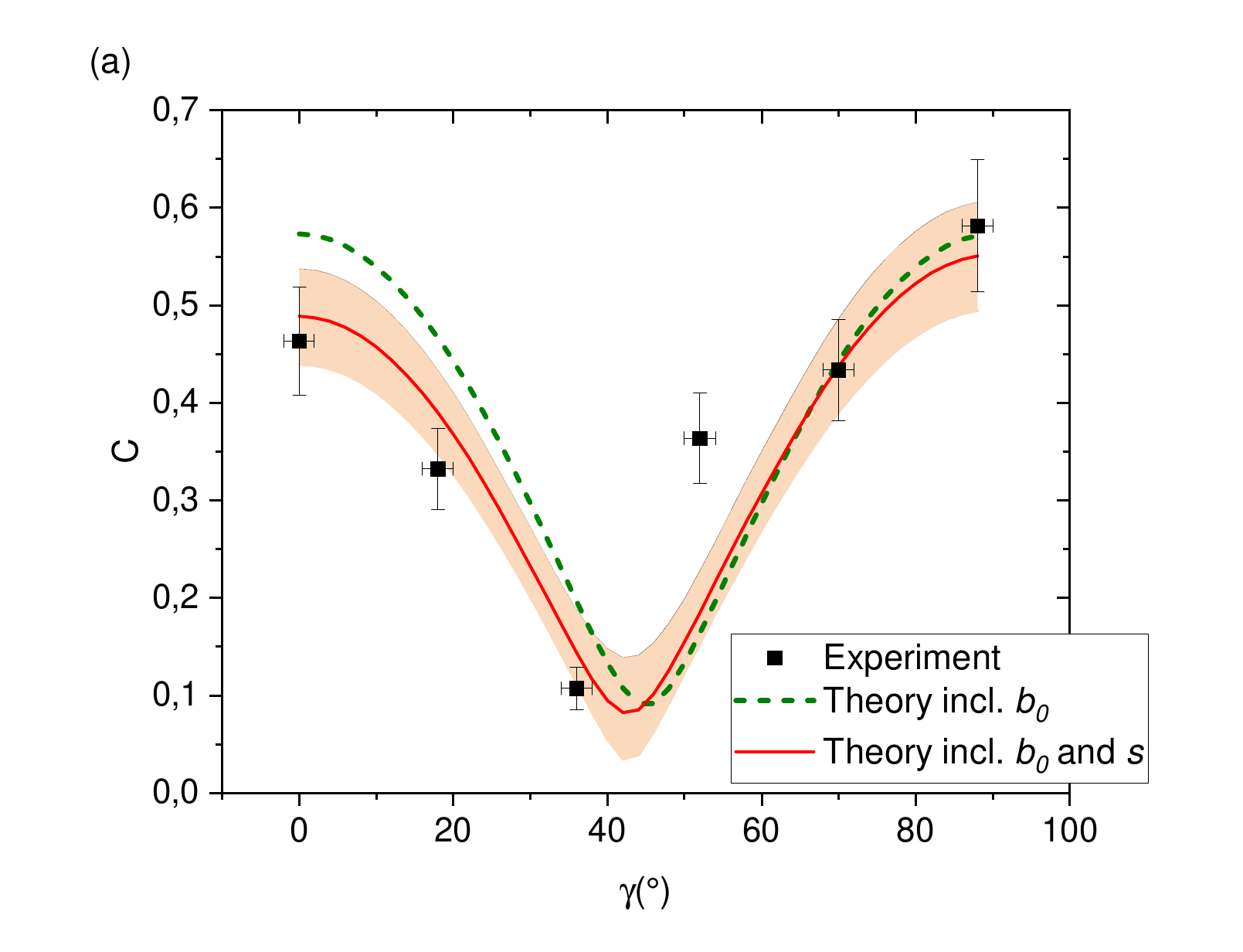}	
	\includegraphics[width=0.45\textwidth]{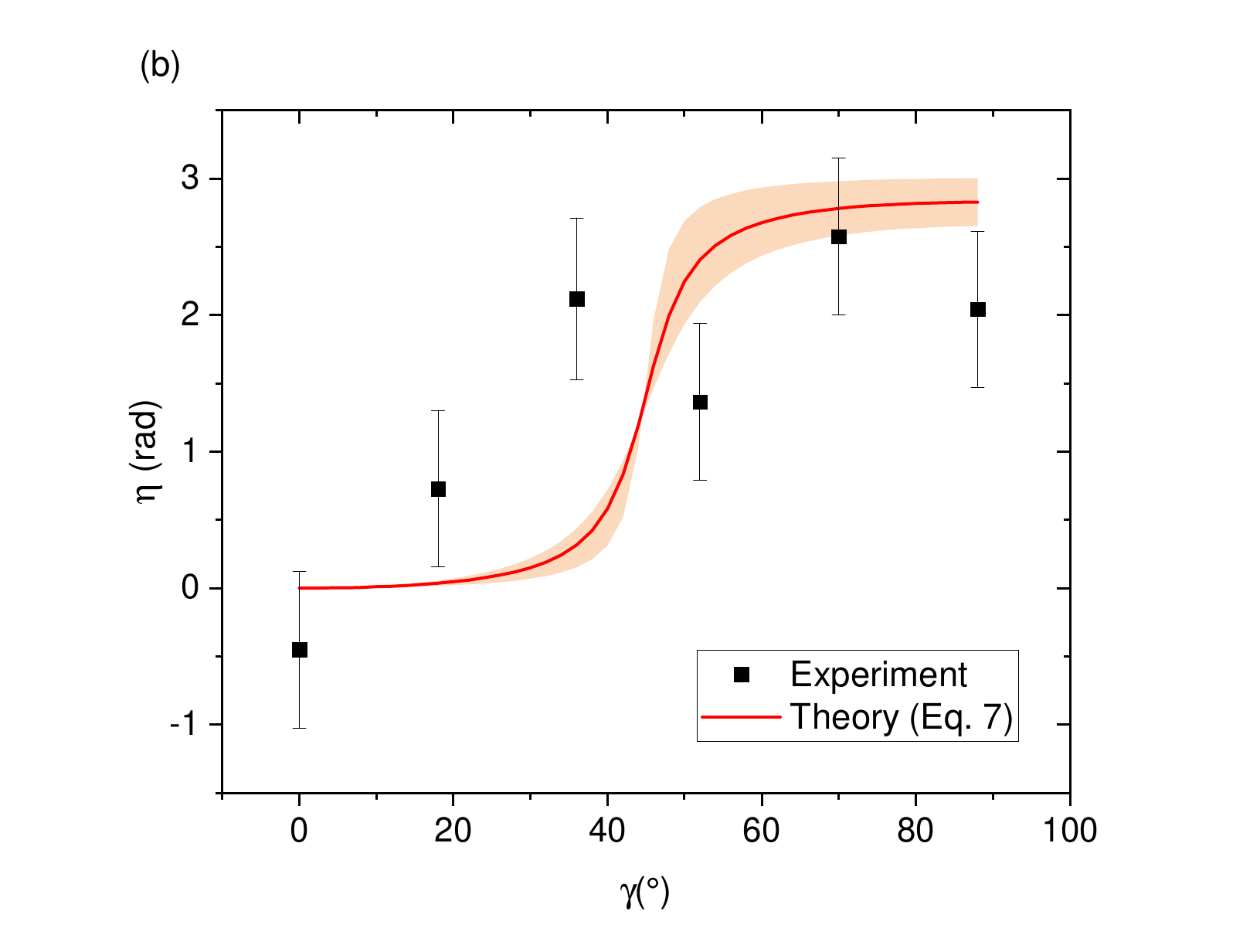}	
	\caption{(a) Contrast derived from the experimental fringes (black squares), from the theoretical prediction~\eqref{eq:C} once the correction due to the optical thickness of the atomic cloud  ($b_0\approx0.4$) is included (dashed curve), plus the correction due to the finite saturation parameter of the incoming beam ($s = 0.07$)(continuous curve). The red shaded area corresponds to the confidence interval, which accounts for the uncertainty on the measurement of $\xi$ and on the fitted rescaling factor to account for $b_0$. (b) Pancharatnam-Berry phase from the experimental fringes displacement (black dots) and from the theoretical prediction~(\ref{eq:phase}). The red shaded area corresponds to the confidence interval associated with the uncertainty on the measurement of $\xi$}. 
    \label{fig:contrast}
\end{figure}

 With our experimental apparatus, we measure the scattered light profile at the CCD camera for different angles of the waveplate (i.e. for different values of $\gamma$). After over 1000 realizations of the experimental sequence described above, we obtain the light scattered as a function of $\theta$, as shown in Fig.~\ref{fig:fringes} (for details on the data analysis and fitting, see \cite{SM}). The contrast $\mathcal{C}$ extracted from those fringes is presented in Fig.~\ref{fig:contrast}(a), and the minimum observed at $\gamma\approx \pi/4$ is  consistent with the prediction from Eq.~\eqref{eq:C}. Note that a better agreement with the experimental data is reached when two effects are accounted for, see plain red curve. First, the finite optical depth of the cloud is responsible for a larger signal from the background-building path~\ref{path:iii} as compared to the others, which in turn reduces the fringes contrast \cite{Moriya16}. Second, the finite saturation parameter $s$ of the laser broadens the spectrum of the light scattered by the atoms, which in turn reduces slightly the contrast of the interference fringes \cite{Dias2021}.  The detailed modelling of these effects, and the associated fitting model, are described in~\cite{SM}.

{\em Pancharatnam–Berry phase \& Geometrical interpretation.---} It is clear from the fringes at Fig.~\ref{fig:fringes} that, as the axis of P is rotated (i.e. as $\gamma$ is varied), the fringes suffer a reduction in contrast yet they are shifted at the same time. Indeed, if we now consider the phase $\delta$ from the scalar product between the polarizations,
\begin{equation}
\langle\hat\epsilon_{i},\hat\epsilon_{ii}\rangle=\mathcal{C}e^{-i\delta},\label{eq:Cc}
\end{equation}
and incorporate the expression in Eq.~\eqref{eq:IntTheta}, then it becomes clear that this phase actually translates into an angular displacement for the fringes. This shift corresponds to the Pancharatnam-Berry phase~\cite{Pancharatnam1956,Berry1987}, and its geometrical interpretation on the Poincaré sphere is presented in Fig.~\ref{fig:physics}: Starting from the injected polarization $\hat\epsilon_0=\hat{x}$, the light is split into paths~\ref{path:i} and ~\ref{path:ii}, with its polarization turned into $\hat\epsilon_i$ and $\hat\epsilon_{ii}$, respectively, before being made to interfere. The solid angle $\Omega$ determined by the geodesic triangle $(\hat{\epsilon}_0\hat\epsilon_{i}\hat\epsilon_{ii})$ on the Poincaré sphere corresponds to twice the phase difference between the polarizations $\hat\epsilon_{i}$ and $\hat\epsilon_{ii}$~\cite{Pancharatnam1956}, that is, $\delta=-\Omega/2$.

From scalar product~\eqref{eq:Cc}, we obtain the following equation for the geometrical phase:
\begin{equation}
    \tan\delta = \frac{\sin\xi\tan^2\gamma}{1+\cos\xi\tan^2\gamma}.
    \label{eq:phase}
\end{equation}
The measurement of the Pancharatnam-Berry phase presents a fair agreement with the theoretical prediction, see Fig.~\ref{fig:contrast}(b). The substantial error bars for the experimental data of Fig.~\ref{fig:contrast}(b) stem from the uncertainty on the determination of the center of the envelope $\theta_0$. Finally, we note that the contrast also has a geometrical interpretation: It corresponds to $\mathcal{C}=\cos(\zeta/2)$, where $\zeta$ is the geodesic distance between the polarizations $\hat\epsilon_{i}$ and $\hat\epsilon_{ii})$ on the Poincaré sphere~\cite{Pancharatnam1956}.

{\em Conclusions.---} We have explored the breaking of optical path reciprocity in a mirror-assisted coherent backscattering setup. The fringes emerging from the light scattered by a disordered cold atomic cloud have their contrast reduced as the optical elements used stop commuting and the two interfering paths become nonreciprocal. The measurement of the contrast and angular displacement of the fringes finds their geometrical interpretation on the Poincaré sphere, encapsulated in, respectively, the geodesic distance between the two polarizations and their phase difference, as originally proposed by Pancharatnam and Berry to provide a measure of the difference between polarizations. 
 
In full generality, these results explore the vectorial nature of light to control the reciprocity of two interfering paths, despite the persisting spectral and spatial coherence. This breaking of reciprocity in this setup mimicks what happens in multiple scattering regime coherence phenomena, such as the CBS and Anderson localization, when time-reversal symmetry is broken \cite{Micklitz2015, Muller2015, Hainaut2018} via e.g. the presence of multilevel Zeeman electronic structure in atoms \cite{Labeyrie1999, Muller2001, Labeyrie2002} or depolarization due to the near-field scattered electric field of close atoms in high densities \cite{Skipetrov2014,Bellando2014}. Interestingly, while the application of a magnetic field can break the time-reversal reciprocity in some cases, it may also restore strong interference phenomena based on the reciprocity of paths, such as for CBS \cite{Sigwarth2004} or Anderson localization \cite{Skipetrov2015}, or as in our setup when the contrast is restored for specific angles of the half waveplate.

Since our disordered sample is made of saturable atoms subjected to classical (thermal) and quantum (spontaneous emission) decoherence \cite{Lassegues2023}, as well as to collective \cite{Guerin2016, Cidrim2020} or many-body effects \cite{Urban2009}, the role of reciprocity for the coherent light emission is usually hidden within the complexity of the system, but could be made explicit, and even controlled, through strategies such as those detailed here, with applications for example to random lasers \cite{Wiersma2008}. Finally, for atoms interacting with a Fabry-Perot cavity, several reflection paths interfere to produce the cavity output; a small nonreciprocal operation, such as a Faraday rotation, on each roundtrip within the cavity could drastically change the spectrum of the system and alter the lasing properties of extended-cavity diode lasers, with a potential to unveil rich new physics at the different cavity-matter interaction regimes.

\vspace{10pt}

\begin{acknowledgments}
 {\em Acknowledgements.---}  R.C.T., R.B. and Ph.W.C. acknowledge funding from Fundação de Amparo à Pesquisa do Estado de São Paulo (FAPESP) through Grants Nos. 2018/23873-3 and 2022/00261-8; R.C.T. and R.B. acknowledge funding from FAPESP through Grants No. 2018/15554-5 and 2023/03300-7; R.C.T., Ph.W.C acknowledge funding from FAPESP through Grant No. 2013/04162-5; Ph.W.C. acknowledges support from the project CAPES-COFECUB Ph879-17/CAPES 88887.130197/2017-01; P.G. S.D., R.K., M.H., R.C.T., R.B. and Ph.W.C. acknowledge funding from ANR + FAPESP (projectQuaCor ANR19-CE47-0014-01/ FAPESP 2019/13143-0) and project STIC-AmSud (Ph879-17/CAPES 88887.521971/2020-00) and CAPES-COFECUB (CAPES 88887.711967/2022-00); R.K. and M.H. acknowledge funding from French National Research Agency (ANR) (project PACE-IN ANR19-QUAN-003-01), from European Unions Horizon 2020 research and innovation programme in the framework of Marie Skodowska-Curie HALT project (grant agreement No 823937), and finantial support of the Doeblin Federation; R.K. acknowledges support from the European project ANDLICA, ERC Advanced Grant Agreement No. 832219; M.H. and R.B. acknowledge support by the French government, through the UCAJ.E.D.I. Investments in the Future project managed by the ANR (ANR-15-IDEX- 01).
\end{acknowledgments}
 
\bibliography{bibli}

\pagebreak

\clearpage

\onecolumngrid
\begin{center}
\textbf{\large Supplemental material: Breaking of reciprocity and the Pancharatnam-Berry phase in a disordered cold atom cloud}\\[.2cm]

\vskip0.5\baselineskip{P. H. N. Magnani,$^{1}$ P. G. S. Dias,$^1$ M. Frometa,$^2$ M. A. Martins,$^1$ N. Piovella,$^3$ R. Kaiser,$^4$ P. Courteille,$^2$ M. Hugbart,$^4$ R. Bachelard,$^{1,4}$ and R. C. Teixeira$^1$}
\vskip0.5\baselineskip{{\em$^{1}$Departamento de Física, Universidade Federal de São Carlos,\\ Rodovia Washington Luís, km 235 - SP-310, 13565-905 São Carlos, SP, Brazil}\\
{\em $^2$ Instituto de F\'isica de S\~{a}o Carlos, Universidade de S\~{a}o Paulo - 13566-590 S\~{a}o Carlos, SP, Brazil}\\
{\em $^{3}$Dipartimento di Fisica "Aldo Pontremoli", Universit\`{a} degli Studi di Milano, Via Celoria 16, I-20133 Milano, Italy \&
INFN Sezione di Milano, Via Celoria 16, I-20133 Milano, Italy}\\
{\em $^{4}$Universit\'e C\^ote d'Azur, CNRS, INPHYNI, France}}
\end{center}

\twocolumngrid

\setcounter{equation}{0}
\setcounter{figure}{0}
\setcounter{table}{0}
\setcounter{page}{1}
\makeatletter
\renewcommand{\theequation}{S\arabic{equation}}
\renewcommand{\thefigure}{S\arabic{figure}}

\section{Fringes theoretical description}

\subsection{Intensity spatial modulation}

Let us consider the scattering of an incident beam with linear polarization $\hat\epsilon_0=\hat{x}$ and wave vector $\mathbf{k}_0 = k(\sin\theta_0\hat{y}-\cos\theta_0\hat{z})$ by $N$ atoms at positions
$\mathbf{r}_j=x_j\hat{x}+y_j\hat{y}+z_j\hat{z}$, in the presence of a $45^\circ$ birefringent mirror M in the plane $(y,z)$, where $\theta_0$ is the (small) angle of incidence with respect to the $z$-axis. An half-wave plate P is inserted between the atomic sample and the mirror, which rotates the polarization of the incoming beam by an angle $\gamma$ in the plane $(x,y)$. The light scattered by each atom is detected in the far field along the direction of the wave vector $\mathbf{k}=k(\sin\theta\hat{y}+\cos\theta\hat{z})$, which for $\theta\approx\theta_0$ is almost anti-parallel to the wave vector $\mathbf{k}_0$ of the incident beam. This backscattered field is the sum of four contributions, given by the four paths described in the main text (see Fig.\ref{fig:setup}).

Supposing initially $s \ll 1$, we can sum independently the four contributions to the total scattered field, given by Eq.~\eqref{eq:E_main}, and one gets the total intensity scattered by a single atom,

\begin{eqnarray}
&&I_s(\mathbf{r}_j)=\frac{I_{a}}{4}\left\|
e^{ik(\cos\theta+\cos\theta_0)z_j}\hat x+e^{ik(\cos\theta-\cos\theta_0)z_j}\hat{\epsilon}_{i}\right.\nonumber\\
&&+\left. e^{-ik(\cos\theta-\cos\theta_0)z_j}\hat{\epsilon}_{ii}
+e^{-ik(\cos\theta+\cos\theta_0)z_j}\hat x
\right\|^2, \label{I1}
\end{eqnarray}
where
\begin{eqnarray}
    \hat{\epsilon}_{i}&=& (\hat{\mathcal{P}}\circ \hat{\mathcal{M}})\hat x
=\cos\gamma \hat x+\sin\gamma\hat y,\label{eq:MP}
\\ \hat{\epsilon}_{ii}&=& (\hat{\mathcal{M}}\circ \hat{\mathcal{P}})\hat x
=\cos\gamma \hat x+e^{i\xi}\sin\gamma\hat y,\label{eq:PM}
\end{eqnarray}
and $I_{a} = 4 E_s^2/(2 \mu_0 c)$ the intensity we would get from the incoherent sum of the single-atom scattered intensity at each of the four paths, with $\mu_0$ the vacuum magnetic permeability and $c$ the light speed in vacuum. 
As for the contribution of path (iv) [last term in Eq.(\ref{I1})], both the incident and scattered beams have been reflected by M through both the blue and green paths of Fig.\ref{fig:setup}, so that
\begin{eqnarray}
(\hat{\mathcal{M}}\circ \hat{\mathcal{P}})\circ(\hat{\mathcal{P}}\circ \hat{\mathcal{M}})
\hat x=\hat{\mathcal{M}}^2\hat x =\hat x,
\label{eq:MPPM}
\end{eqnarray}
since $\hat{\mathcal{P}}^2=1$.
Hence, the polarization of the scattered incident beam remains unchanged due to the double passage through P and M.
 
By summing over the random positions $z_j$ of the atoms, assuming $\theta-\theta_0\ll\theta_0$, $b_0 \ll 1$ and a Gaussian atomic distribution, we find that the total scattered intensity $I_{\text{out}}$ is spatially modulated as follows:
\begin{eqnarray}
&&I_{\textrm{out}}=N I_a\bigg\{1+e^{-2(k\theta_0 R)^2(\theta-\theta_0)^2}\times \nonumber\\
&&\frac{1}{2}\Big[
\cos^2\gamma\cos[2\theta_0kh(\theta-\theta_0)]
+ \\
&& \sin^2\gamma\cos[2\theta_0kh(\theta-\theta_0)-\xi]\bigg]
\Big\},\nonumber
\label{eq:Is}
\end{eqnarray}
since $\langle\cos[2k(\cos\theta+\cos\theta_0)z]\rangle_j=0$ and
\begin{eqnarray}
& \langle\cos[2k(\cos\theta-&\cos\theta_0)z]\rangle =e^{-2(k\theta_0\ R)^2(\theta-\theta_0)^2}\nonumber\\
&&\times \cos[2\theta_0kh(\theta-\theta_0)] .
\end{eqnarray}
These formulas are valid since the two angles $\theta$ and $\theta_0$ are small, and the cloud large compared to the wavelength of the scattered light. Using a polarization analyzer from Shäfter und Kirchhoff (model SK010PA-VIS), we measured the relative phase shift between the $\hat{x}$ and $\hat{y}$ components of the polarization induced by M on the reflected light, to be  $(81 \pm 5)^{\circ}$; this corresponds to the value $\xi = (162 \pm 10)^{\circ}$ since the light is reflected twice by it.

\subsection{Fringes contrast and displacement}

Let us introduce the phase $\phi=2\theta_0kh(\theta-\theta_0)$ to describe the angular fringes. Then the interference fringe term in Eq.~\eqref{eq:Is} can be written as
\begin{equation}
\cos^2\gamma\cos\phi+\sin^2\gamma\cos(\phi-\xi)=\mathcal{C}\cos(\phi-\delta).
\end{equation}
By expanding the cosines, we obtain
\begin{eqnarray*}
(\cos^2\gamma+\cos\xi\sin^2\gamma)\cos\phi &+& \sin\xi\sin^2\gamma\sin\phi\nonumber\\
=\mathcal{C}\cos\delta\cos\phi &+&\mathcal{C}\sin\delta\sin\phi,
\end{eqnarray*}
which leads to the set of equations:
\begin{eqnarray*}
\mathcal{C}\cos\delta&=& \cos^2\gamma+\cos\xi\sin^2\gamma\\
\mathcal{C}\sin\delta&=& \sin\xi\sin^2\gamma
\end{eqnarray*}
By squaring and summing, we obtain the squared contrast
\begin{eqnarray*}
\mathcal{C}^2&=&\cos^4\gamma+\sin^4\gamma+2\cos\xi\sin^2\gamma\cos^2\gamma\nonumber\\
&=& \frac{1}{2}+\frac{1}{2}\cos^2(2\gamma)+\frac{1}{2}\cos\xi\sin^2(2\gamma)\\
&=& 1 +\frac{1}{2}\sin^2(2\gamma)(\cos\xi-1),
\end{eqnarray*}
and thus the contrast
\begin{equation}
\mathcal{C}=\sqrt{1-\sin^2(2\gamma)\sin^2(\xi/2)}.
\end{equation} 
The fringe displacement $\delta$ is determined by
\begin{equation}
\tan\delta=\frac{\sin\xi\tan^2\gamma}{1+\cos\xi\tan^2\gamma}.
\end{equation}

Those two parameters are also exactly found by the scalar product between the two polarization angles, as appearing in Eq.~(\ref{eq:Cc}), such as to recover 

\begin{equation}
    I_\textrm{out}(\theta)=NI_a \left(1+\frac{1}{2} \Re[\langle\hat\epsilon_{i},\hat\epsilon_{ii}\rangle e^{i\phi}]S_{R}(\theta)\right)
\end{equation}

\noindent which corresponds to Eq.~\eqref{eq:IntTheta} of the main text; or still

\begin{eqnarray}
        I_\textrm{out}(\theta)=NI_a \bigg(1+\frac{1}{2} \, \mathcal{C} e^{-2(k\theta_0\ R)^2(\theta-\theta_0)^2}  \\ 
     \times \cos[2\theta_0 kh(\theta-\theta_0) - \delta] \bigg) \ .
\end{eqnarray}

\subsection{Corrections due to $s$ and $b_0$}

For finite saturation parameter, the contrast of the setup becomes a function of the saturation parameter and of the time needed for the light to travel from the atomic cloud to the backscattered mirror and back (i.e., the time needed to perform either the path $A'$ or $B'$ on Fig.~\ref{fig:setup}), because the spectrum of the inelastically scattered light is broadened with respect to the incoming laser spectrum, and the scattered light following path $B'$ gets dephased with respect to the scattered light following path $B$. Following \cite{Dias2021}, we write the saturation parameter $s_1$ seen by an individual atom at position $z$:

\begin{equation*}
s_1(z) = 2s \left[1 + \cos \gamma \, \cos \left(2 k \cos \theta_0 z \right)\right] \ .
\end{equation*}

For $\gamma = 0$, the polarization of the incoming and backreflected laser light is the same, and an intensity grating is created; while for $\gamma = \pi/2$, the polarization is crossed and all atoms see the same intensity. The first-order temporal correlation of the light scattered by this atom in the rotating reference frame of the incoming laser light, $\tilde{g}^{(1)}_{z} (\tau_c)$, is given by

\begin{eqnarray}
 &&\tilde{g}^{(1)}_{z} (\tau_c) = \frac{1}{1 + s_1(z)} + \frac{1}{2} \left[\text{e}^{-\Gamma \tau_c/2}  + \right. \nonumber\\
 &&\frac{s_1(z) - 1}{s_1(z) + 1} \cos \left(\Omega_M(z)\, \tau_c \right) \text{e}^{-3 \Gamma \tau_c/4} + \\
 && \left.\frac{\Gamma}{4\Omega_M(z)} \frac{5s_1(z) - 1}{s_1(z) + 1} \sin \left(\Omega_M(z) \, \tau_c \right) \text{e}^{-3 \Gamma \tau_c/4} \right]  \nonumber 
 \label{eq:g1_z}
\end{eqnarray}

\noindent with $\Omega_M (z) = \Gamma \sqrt{\tfrac{s_1(z)}{2} - \tfrac{1}{16}}$. The intensity scattered by a single atom, Eq.~(\ref{I1}) is accordingly modified, and after some algebra, it gives

\begin{eqnarray}
&& I_s(\mathbf{r}_j)=4 I_{1} \Bigg[2 + \tilde{g}^{(1)}_z (\tau_c) \, \cos \left(2 k z \cos \theta \right) \times \label{eq:I1_s}\\
&& \frac{2 \cos \gamma \, + \, \cos \left(2 k z \cos \theta_0 \right) + \mathcal{C}\cos \left(2 k z \cos \theta_0 + \delta \right)}{1 \, + \, \cos \gamma \, \cos \left(2 k z \cos \theta_0 \right)} \Bigg] \nonumber
\end{eqnarray}

We note at this point that this last expression is different from what is found in \cite{Dias2021}, since there is no transformation $\hat{\mathcal{P}}$ included there; we can then find the expression of Ref. \cite{Dias2021} for the emission of a single atom by making $\xi = 0$ at Eq.~(\ref{eq:I1_s}). Upon integrating the emission profile \ref{eq:I1_s} over a Gaussian atomic distribution, we obtain an intensity profile given by

\begin{eqnarray}
&&I_\textrm{out}=N I_a\left\{1+\frac{1}{2}\,\mathcal{C}_s (s,\tau_c,\gamma, \xi) \,e^{-2(k\theta_0 R)^2(\theta-\theta_0)^2}\times \right.\nonumber\\
&&\left. \cos[2\theta_0kh(\theta-\theta_0) - \delta]\right\} \ , 
\end{eqnarray}

\noindent with $\mathcal{C}_s (s,\tau_c,\gamma, \xi)$ the contrast corrected by a finite $s$, found by numerical integration of Eq.~(\ref{eq:I1_s}) over the Gaussian atomic distribution; $\tau_c = 4.0~$ns is the light dephasing time for travelling to the backscattered mirror and back. For additional correction on the finite optical density of the cloud, we note that it causes an attenuation of the field of each path, yet it is less for path~\ref{path:iii} than for paths~\ref{path:i} and~\ref{path:ii}. Because the path ~\ref{path:iii} contributes to the background signal, at first order it leads to a reduction of the contrast by a factor $1-b_0$~\cite{Moriya16}. Since $s$ is small, we keep the correction from the linear scattering regime \cite{Moriya16} and write as final expression for fitting the fringes

\begin{eqnarray}
&&I_\textrm{out}=NI_a\bigg\{1+\frac{1}{2}\,\left(1 - b_0\right)\,\mathcal{C}_s (s,\tau_c,\gamma, \xi) \times  \nonumber\\
&& \,e^{-2(k\theta_0 R)^2(\theta-\theta_0)^2} \cos[2\theta_0kh(\theta-\theta_0) - \delta]\bigg\} \ . \label{eq:final}
\end{eqnarray}

\section{Fringe Measurements}

A CCD camera is used to collect the backscattered light (path $B$), which is separated by $2 \theta_0 = 8.6^{\circ}$ from the direction of the reflected beam, in order to minimize the effect of stray light. A lens conjugates the far-field scattering profile at its focal plane, where the CCD camera is placed. For each set of parameters, the average over at least 1000 realizations is performed, with three images extracted at each realization: one in presence of the atoms and the laser beam, one in presence only of the laser beam, and one without atoms and laser beam. We then obtain three far-field intensity profiles, $I_1(\theta,\varphi)$, $I_2(\theta,\varphi)$ and $I_\varnothing(\theta,\varphi)$, from the average of these images, respectively, for all realizations. The profile $I_\varnothing$ contains only spurious effects such as noise from the camera and other sources of stray light apart from the incident laser beam, which are also present in the profiles $I_1$ and $I_2$. We thus subtract it from the picture with atoms and laser, and from that with the laser only, yielding $I_a = I_1 - I_\varnothing$ and $I_l = I_2 - I_\varnothing$. The profile $I_a$ now contains the light scattered by the atoms we want to extract, as well as the stray light created by the reflected laser beam due to unavoidable light scattered by the different optical elements. While the stray light presents on $I_a$ is attenuated due to the absorption of the laser beam by the atoms, the profile $I_l$ contains only the stray light, yet not attenuated. Outside the region where the fringes appear, the light scattered by the atoms has a constant value, while the spatial distribution of the laser stray light is not isotropic but centered around the direction of the reflected laser beam. Hence, in order to extract only the light scattered by the atoms, we perform a fit of the intensity profiles outside the fringe region with the expression
\begin{equation}
I_a(\theta,\varphi) = F + T \,I_l(\theta,\varphi), 
\end{equation}
where $F$ and $T$ are positive parameters which account for the background fluorescence and the transmission coefficient of the reflected beam through the atomic cloud, respectively. From this fit, we obtain the 2D intensity profile $I_\textrm{scat}(\theta,\varphi)$ of the light scattered by the atoms in the fringe region, according to the equation 
\begin{equation}
I_\textrm{scat}(\theta,\varphi) = I_a(\theta,\varphi) - T \,I_l(\theta,\varphi). 
\end{equation}

The 2D profile is then integrated along the azimuthal angle to give a 1D profile ($I_\textrm{out}$ of Eq.~\eqref{eq:IntTheta}), that allows us to extract all relevant parameters, the contrast and the phase shift. The 1D experimental fringes used for this article are shown in Fig.~\ref{fig:fringes}.

\section{Fitting protocol} 

The contrast $\mathcal{C}$ and fringe displacement $\delta$ are extracted from the experimental signal by fitting the fringes shown in Fig.~\ref{fig:fringes} using Eq.~\eqref{eq:IntTheta} with $I_a$, $\mathcal{C}$, $\delta$, $\theta_0$, $h$ and $R$ as free parameters (see continuous lines in Fig.~\ref{fig:fringes}). While $\mathcal{C}$, $\delta$ and $I_a$ are let free for each value of $\gamma$, the parameters $\theta_0$, $h$ and $R$ are shared by all curves, which is justified by the stability of the atomic cloud and the alignment of the incoming laser beam through the optical path to the CCD camera (those parameters depend only on the light propagation direction and on the atomic cloud geometry). However, as shown in Fig.~\ref{fig:fringes}, the experimental data present a slight lower value, as well as a pronounced asymmetry, which can be attributed to the cloud finite optical depth $b_0$ and to the pump finite saturation parameter $s$, respectively, as discussed in the section above. The behavior of the contrast $\mathcal{C}_s (s,\tau_c,\gamma, \xi)$, corrected for the finite $s$, is fitted to the experimental points of Fig.~\ref{fig:contrast}(a), with only free parameter the multiplicative factor $1-b_0$. We find in this way the value $b_0 = 0.43$, in good agreement with the value $0.4(1)$ directly measured by absorption imaging.

\end{document}